\documentclass[3p,times]{elsarticle}

\usepackage{ecrc}

\volume{00}

\firstpage{1}

\journalname{Annals of Physics}

\runauth{A. N. Rubtsov et al.}

\jid{procs}

\jnltitlelogo{Annals of Physics}

\CopyrightLine{2011}{Published by Elsevier Ltd.}

\usepackage{amssymb}

\usepackage[figuresright]{rotating}

\begin{document}

\begin{frontmatter}



\dochead{}

\title{Dual boson approach to collective excitations in correlated fermionic systems}


\author{A.~N.~Rubtsov}
\address{Department of Physics, M.V. Lomonosov Moscow State University,
119991 Moscow, Russia}
\ead{ar@ct-qmc.org}
\author{M.~I.~Katsnelson}
\address{Radboud University Nijmegen, Institute for Molecules
and Materials, 6525AJ Nijmegen, The Netherlands}
\author{A.~I.~Lichtenstein}
\address{Institute of theoretical Physics, University of Hamburg, 20355
Hamburg, Germany}

\begin{abstract}
We develop a general theory of a boson
decomposition for both local and non-local interactions in lattice fermion models which
allows us to describe fermionic degrees of freedom and collective charge
and spin excitations on equal footing. An efficient perturbation theory in
the interaction of the fermionic and the bosonic degrees of freedom is
constructed in so-called dual variables in the path-integral formalism.
This theory takes into account all local correlations of
fermions and collective bosonic modes and interpolates
between itinerant and localized regimes of electrons in solids.
The zero-order approximation of this theory corresponds to
extended dynamical mean-field theory (EDMFT), a regular way to calculate
nonlocal corrections to EDMFT is provided. It is shown that dual ladder summation
gives a conserving approximation beyond EDMFT. The method is especially
suitable for consideration of collective magnetic and charge excitations and
allows to calculate their renormalization with respect to ``bare'' RPA-like
characteristics. General expression for the plasmonic dispersion in correlated media is obtained.
As an illustration it is shown that effective superexchange
interactions in the half-filled Hubbard model can be derived within the
dual-ladder approximation.
\end{abstract}


\end{frontmatter}


\section{Introduction}
\label{Introduction}

A self-consistent description of electrons, magnons, phonons and other
fermionic and bosonic degrees of freedom in solids is still a challenging
problem for strongly correlated systems where the simplest approximations are
usually not sufficient. One of the classical examples is the theory of
itinerant electron magnetism, where the d-electrons form both the
correlated energy bands as well as the quasi-localized magnetic moments and
the magnon spectrum \cite{Moriya}. In this problem the conventional random phase
approximation (RPA) gives a wrong description of the finite temperature effects
whereas the proper corrections to RPA can be successfully calculated only in
a very close vicinity of the Stoner instability point \cite
{Moriya,Dzyalos_Kondrat}. The main difficulty is the coexistence of
localized (atomic-like) and itinerant features in the behavior of
d-electrons in solids \cite{vons_kats_tref1993a,vons_kats_tref1993b,ourPRL_Fe_Ni2001}.
Recent progress in the theory of correlated electron systems is related to the development of
Dynamical Mean Field Theory (DMFT) \cite{Georges_RMP}, which maps the
problem of fermions with local interactions on a crystal lattice onto
a self-consistent effective impurity problem.
The latter can be solved very efficiently within Continuous Time Quantum
Monte Carlo (CT-QMC) \cite{Gull_RMP_CTQMC} or exact diagonalization (ED)\cite{Georges_RMP} schemes.
As a result, both atomic-like
features such as multiplet formation and band dispersion can be taken into
account simultaneously \cite{Umat_Pu,CoinCu}. The combination of the DMFT
approach with realistic electronic structure calculations \cite
{Kotliar_RMP,LK1998}, in particular, for transition metal systems \cite
{ourPRL_Fe_Ni2001,ourRMP} opens a new way for the description of electronic
degrees of freedom in solids. A generalization of the DMFT approach to
non-local interactions and consistent description of collective bosonic
excitations \cite{Kotliar_RMP,Chitra_Kotliar,Sun_Kotliar_EDMFT,optics} boost first
realistic applications of a so-called GW+DMFT approach \cite{Biermann_GW+DMFT}.

Here we construct  a theory, that is capable to describe systems with
strongly correlated electrons interacting with collective bosonic modes.
These modes can appear explicitly in an original Hamiltonian (for instance,
phonons) or arise from electron-electron interactions (like plasmons or
magnons). We assume that the Bose excitations cannot be considered as an
ideal Bose gas and boson-boson correlations are important. Among numerous
examples of such systems are the cases of weak itinerant ferromagnets \cite{Dzyalos_Kondrat} and, more generally,
any materials near quantum phase transition points \cite{sachdev}.

The Feynmann diagrammatic technique became the most popular tool
in the quantum many body theory as a compact and elegant representation of
a perturbation expansion in powers of interaction \cite{AGD,Mahan}.
The main advantage of the diagrammatic approach is an opportunity to sum up
infinite series of relevant diagrams going beyond the formal perturbation
expansion. The mainstream paradigm is that summing up more sophisticated series
expands the area of applicability.
In some cases leading series of diagrams can be separated and summed up
with the use of small parameters which are not directly related to the interaction
strength. Important examples are the Gell-Mann Bruekner theory of the high-density electron gas \cite{GB},
the gaseous approximations of low density Fermi \cite{Galitskii}  and Bose \cite{Belyaev} systems,
the $\varepsilon$-expansion in theory of critical phenomena \cite{Wilson_Kogut}, and the $1/N$ expansion
in the theory of magnetic impurities \cite{Bickers}. Historically, the DMFT was introduced as a leading
order approximation in a formal parameter $1/d$, where $d$ is the dimensionality of the space.
DMFT is exact in the limit of $d \rightarrow \infty$, at the same time attempts to construct
a regular expansion in   $1/d$ turn out to be not practical \cite{Shiller}. Nevertheless, DMFT
appears to be quite a good approximation for many three- and even two-dimensional systems
\cite{Georges_RMP,Kotliar_RMP,ourRMP}, since it provides reasonable interpolations between
non-interacting and atomic limits for any dimensions. In terms of the diagrammatic approach the DMFT
corresponds to a summation of all local in space contributions to the self-energy \cite{Georges_RMP}.

Recently we developed the dual fermion scheme \cite{Rubtsov_DF,Rubtsov_big_PRB} which
introduces new variables in a path integral generating an alternative diagram technique such that
the bare propagators (free dual fermions) correspond to DMFT. This means that the main local
physics is taken into account by the change of variables while the new diagrams describe only non-local
corrections to the self-energy. Interaction terms in the dual variables look more complicated than
the original ones: the interactions become retarded and contain not only pair terms but also multi-fermion
contributions. However, an advantage of dual fermion transformation is a faster convergence of the diagram
series. For example, a simple ladder summation gives quite accurate description for the single band Hubbard
model in strongly correlated regime \cite{Hafermann_Ladder}. Thus, the dual fermion approach is
an example of an alternative strategy in the quantum many-body theory, when a proper choice of variables
allows one to consider a simpler series of diagrams. Another example of such strategy is the use of auxiliary
fermions and bosons in the Hubbard model which allows one to describe the subtle effect of the growth of
the effective mass near metal-insulator transitions, already on the level of the mean-field  \cite{Kotliar-Ruckenstein}.

The consideration of collective (bosonic) degrees of freedom in systems of interacting fermions
within the conventional technique necessarily involves  summations of sophisticated diagrams.
The dispersion law of bosonic excitations (e.g., magnons or plasmons) is determined by the poles
of the corresponding susceptibilities (the two particle fermionic Green's functions) and the minimal
approximations which give rise to these poles, even in the case of a weak interaction corresponding to the ladder
summation (random phase approximation, RPA \cite{AGD,Bohm_Pines,Khodel_RPA}).
In order to take into account magnon-magnon interactions
one needs to consider more complicated, parquet-like diagrams \cite{Dzyalos_Kondrat,Irkhin_Katanin_Katsnelson}.
For strongly correlated systems one can consider the same procedure within the
dual fermion approach. Up to now this is only done at the level of ladder summations \cite{Hafermann_Ladder}.
As we pointed out, the above considerations of complicated diagrams can be replaced by a proper choice
of variables in the path integral. Following this strategy we propose an extension of dual fermion scheme
explicitly adding bosonic variables. It is important to note that the present theory is not a direct extension of
the dual-fermion approach \cite{Rubtsov_DF}. We will see that, even for the
Hubbard-like models with on-site interactions only, it gives a different
result because of the account of collective modes (magnons).

A simultaneous description of fermionic and bosonic correlations on the
equal footing was started within the Extended Dynamical Mean Field (EDMFT)
framework \cite{Chitra_Kotliar,Sun_Kotliar_EDMFT}. The strong-coupling
generalization of  EDMFT \cite{Stanescu_Kotliar} gives a useful extension
of the Hubbard-I like perturbation expansion for lattice models \cite
{Pairault_Senechal_Tremblay}. EDMFT properly includes the physics of
\textit{local} fermionic and bosonic correlations. At the same time, a dispersion
of collective excitations within EDMFT is determined by
the non-local part of the interactions {\it only}. In particular, to describe the dispersive magnons
within EDMFT one needs to  explicitly add direct Heisenberg exchange
interactions to the Hamiltonian.
Obviously, this is not sufficient since the {\it indirect} exchange interactions
(superexchange, double-exchange, RKKY, etc.) contribute as well to the the magnon dispersion.
Moreover, typically, they give the main contribution.

In the present paper we construct a formally exact diagram technique where
the bare fermionic and bosonic propagators correspond to EDMFT, whereas
the indirect exchange is given by diagrammatic corrections. Therefore, the
direct and indirect interactions enter the magnon dispersion law on equal footing.

\section{Definitions}

We consider a lattice fermionic problem with the intersite hopping and
generic non-local interactions. For simplicity, we present only formulae for
the single-band periodic lattice, but a generalization of all expressions to a
multiorbital case is straightforward. The action is
\begin{equation}  \label{Lattice}
S=\sum_r S_{at}[c^\dag_r, c_r]+\sum_{r,R\neq 0, \omega, \sigma}
\varepsilon_R c^\dag_{r \omega \sigma} c_{r+R \omega \sigma} +\sum_{r,R\neq
0, \Omega} V_{R \Omega} \rho_{r \Omega}^*
\rho_{r+R \Omega}
\end{equation}
Here $c^\dag,\,c$ are Grassmann variables, the indices $r, R$ run over
the lattice sites and translations, $k$ stands for the wave-vector, $\tau$ is
the imaginary time,
$\omega=2\pi (n+1/2) / \beta $ and $\Omega=2\pi n / \beta $ are, respectively, the fermionic and bosonic
Matsubara frequencies ($n$ is integer and $\beta$ is the inverse temperature), and $\sigma$ is the spin index.
Prefactors $\beta^{-1}$ and $N^{-1}$ are implied for sums over frequencies and wavevectors,
respectively  ($N$ is the number of lattice sites).
The ``Bosonic variable'' $\rho$ in general is a vector quantity,
whose components for single band models can be the charge-density or the spin-density
operators: $\rho_j \equiv \sum_{\sigma \sigma^{\prime}} c^\dag_{\sigma}
s^{j}_{\sigma \sigma^{\prime}} c_{\sigma^{\prime}}-\bar{\rho}$, where $s^j$ with
$j=\{0, x, y, z\}$ are Pauli matrices ($1, \sigma_x, \sigma_y, \sigma_z, $), and the quantity $\bar{\rho}$
is introduced to satisfy the condition $<\rho>=0$.

To simplify the notations, we skip the index $j$ throughout the paper, but
one should bear in mind that relevant components of $\rho$ are to be accounted in a practical calculation.
For example,  if the physics is determined by charge fluctuations only (as may be the case
for the Coulomb interaction or an electron-electron attraction mediated by phonons),
only the density-density term
$V^{00} \rho_0^* \rho_0$ is important in the interaction part $V^{j j^{\prime}}_{R \Omega}$ of Eq.(\ref{Lattice}). The Heisenberg isotropic exchange interaction corresponds
to $V^{xx}=V^{yy}=V^{zz}=J$, etc.
An important case of the Hubbard model corresponds to $V=0$, where the spin degrees of freedom $\rho_{x,y,z}$ do not show up in the bare expression part of Eq.(\ref{Lattice}), but will appear in the following formulation, since the
low-energy behaviour of system is governed by the spin fluctuations.

Within the path integral representation the atomic action $S_{at}$ is given
by:
\begin{equation}  \label{Atom}
S_{at}=-\sum_{\omega\sigma} (i \omega +\mu) c^\dag_{\omega\sigma}
c_{\omega\sigma} + \int_0^\beta U c^\dag_{\uparrow} c_{\uparrow}
c^\dag_{\downarrow} c_{\downarrow} d\tau,
\end{equation}
where $U$ is the local Coulomb interaction.

We are interested in the single-electron Green's function
$G_{r \tau}=-<c_{r \tau} c^\dag_{r=0, \tau=0}>$ and the bosonic correlator $X_{r
\tau}=-<\rho_{r \tau} \rho_{r=0, \tau=0 }^*>$. It is an important point that
an account of the bosonic quantity is fundamental: we suppose that
collective modes are relevant for the physics of the system.

\section{Local approximations}

The general idea of effective-medium approximations is to reduce the lattice
problem (\ref{Lattice}) to the so-called impurity model, describing a single
atom (or small cluster) in a Gaussian environment. For a system with bosonic
and fermionic degrees of freedom, a single-site impurity problem is
described by the action
\begin{equation}  \label{Impurity}
S_{imp}=S_{at}+\sum_\omega \Delta_\omega c^\dag_\omega c_\omega+\sum_\Omega
\Lambda_\Omega \rho^*_\Omega \rho_\Omega .
\end{equation}

Obtaining a numerical or analytical solution for the impurity problem
is supposed to be much simpler. We later assume that its Green's functions: $%
g_{\tau}=-<c_{\tau} c^\dag_{\tau=0}>_{imp}$ and $\chi_{\tau}=-<\rho_{\tau}
\rho_{\tau=0}^*>_{imp}$ are known, as well as higher-order correlators.
In practice, they can be calculated by a CT-QMC scheme \cite{Gull_RMP_CTQMC}
or by exact diagonalization if the dimension of the Hilbert space is not too
large.
Normally $\chi$ is called the full polarizability. We will call it the bosonic Green's function, to
emphasize the formal analogy between the equations for $g$ and $\chi$ in our formalism.

For the so-called local approximations, only the Green's functions of the
impurity model $g_{\omega}, \chi_\Omega$ are taken into account. In a
nutshell, the initial problem (\ref{Lattice}) is replaced by an auxiliary
lattice with Gaussian nodes; those nodes are characterized by the Green's
functions $g_{\omega}, \chi_\Omega$. Such a consideration gives us the well-known
expressions for lattice Green's functions \cite{Sun_Kotliar_EDMFT}, which
are denoted with calligraphic letters in this paper:
\begin{eqnarray}\label{MFG}
{\cal G}_{\omega k}=\frac{1}{g^{-1}_\omega+\Delta_\omega-\varepsilon_k}   \\
{\cal X}_{\Omega K}=\frac{1}{\chi^{-1}_\Omega+\Lambda_\Omega-V_K} \nonumber
\end{eqnarray}
where $k$ is the wave vector and Fourier transforms of the hopping and
interaction parameters are introduced in the obvious way. One can see that
the fermionic and bosonic self-energies of these expressions are $k$%
-independent, that is, local in space.

The hybridization functions $\Delta$ and $\Lambda$ should be defined in such
a way, that the impurity problem provides us the on-site (local) properties
of the solution for the lattice problem. Such a requirement is  physically quite
reasonable. Also, it is closely related to
Feynman's variational approach in dual fermion variables \cite{Rubtsov_big_PRB}.

Most common are so-called static and dynamic mean-field theories. In the
static approach, the hybridization functions $\Delta $ and $\Lambda $ are
chosen to be frequency independent. Their values are defined from the
equalities for the \textit{static parts} of Green's functions:
\begin{eqnarray}\label{MFcondition}
\sum_\omega g_\omega^{MF}=\sum_{\omega k} {\cal G}_{\omega k}^{MF}, \\ \nonumber
\sum_\Omega \chi_\Omega^{MF}=\sum_{\Omega K} {\cal X}_{\Omega K}^{MF}
\end{eqnarray}

In the dynamical EDMFT approach $\Delta_\omega$ and $\Lambda_\Omega$ are
adjusted so that local parts of the lattice Green's functions coincide with
the corresponding quantities of the impurity problem \textit{for all frequency
(or time) arguments}:
\begin{eqnarray}\label{DMFTcondition}
g_\omega=\sum_k {\cal G}_{\omega k}, \\ \nonumber
\chi_\Omega=\sum_K {\cal X}_{\Omega K}
\end{eqnarray}

It is worthwhile to note that the standard definition of the EDMFT \cite
{Sun_Kotliar_EDMFT} introduces an auxiliary bosonic field called $\phi$, and
the second line of Eq.(\ref{DMFTcondition}) is re-expressed as a condition
for the Green's function $\mathcal{X}_\phi$ of that bosonic field. At the
EDMFT level, the conditions for $\mathcal{X}$ and $\mathcal{X}_\phi$ are
equivalent. We discuss this issue in detail in Appendix and show that
similar (although not the same) bosonic field can be introduced in non-local
approximations beyond the EDMFT.

Some ``mixed'' local approximations can also make sense. For instance, if
the hopping is absent (or negligible), it is natural to consider a scheme
with $\Delta=0$ and to adjust bosonic hybridization only. Contrary, $%
\Lambda=0$ corresponds to the ``purely fermionic'' DMFT scheme \cite
{Georges_RMP}.

\section{Dual transformation}

It is easy to make a connection between the lattice and impurity actions:

\begin{equation}  \label{Lattice_imp}
S=\sum_r S_{imp}[c^\dag_r, c_r]+\sum_{\omega, k, \sigma} \left( \varepsilon_k -
\Delta_{\omega\sigma} \right) c^\dag_{\omega k \sigma} c_{\omega k \sigma}
+\sum_{\Omega, K} \left( V_{\Omega K}-\Lambda_{\Omega} \right) \rho_{\Omega K }^* \rho_{\Omega K}
\end{equation}

Further on, we use a general matrix form of the Hubbard-Stratonovich
transformation to decouple the term $c^\dag_{\omega k \sigma} E_{\omega k
\sigma} c_{\omega k \sigma}$ with $E_{\omega k
\sigma}=\Delta_{\omega\sigma}-\varepsilon_{k}$ into the pair of dual fermionic
Grassmann-fields $f^\dag_{\omega k \sigma}, f_{\omega k \sigma} $ and the
term $\rho^*_{\Omega K j} W^{j j^{\prime}}_{\Omega K} \rho_{\Omega K
j^{\prime}} $ with $W^{j j^{\prime}}_{\Omega K} = \Lambda^{j
j^{\prime}}_{\Omega}-V^{j j^{\prime}}_{\Omega K}$ into the pair of dual
bosonic complex-fields $\eta^*_{\Omega K j}, \eta_{\Omega K j}$ (a matrix
notation is used to shorten the expressions):
\begin{eqnarray}  \label{HS}
\begin{array}{l}
\int D\left[c^\dag, c \right]e^{c^\dag E c} = \int D\left[c^\dag, c \right]
\det\left(\alpha^{-1}_f E \alpha^{-1}_f \right) \int D\left[f^\dag, f \right]
e^{-f^\dag \alpha_f E^{-1} \alpha_f f + f^\dag \alpha_f c + c^\dag \alpha_f f } \\
\\
\int D\left[\rho^*, \rho \right]e^{\rho^* W \rho} = \int D\left[\rho^*, \rho
\right] \det\left(\alpha_b W^{-1} \alpha_b \right) \int D\left[\eta^*, \eta \right]
e^{-\eta^* \alpha_b W^{-1} \alpha_b \eta + \eta^* \alpha_b \rho + \rho^* \alpha_b \eta } ,
\end{array}
\end{eqnarray}
where $\alpha_f, \alpha_b$ are arbitrary, but local in space scaling
matrices.

Note that whereas the decoupling of Grassmanian variables is mathematically
straightforward, more comments are needed on the second line of Eq.(\ref{HS}).
The first comment is that $\rho$ is Hermitian, so that $\rho^*_{-\Omega -K
j}=\rho_{\Omega K j}$, and the integral $D[\rho^*, \rho]$ is taken in fact over
the variables from a half-space. The dual variables inherit this property, $%
\eta^*_{-\Omega -K j}=\eta_{\Omega K j}$.
The second comment is that the matrix $\alpha _{b}W^{-1}\alpha _{b}$ is not
positively defined, and integrals over $D[\eta ^{\ast },\eta ]$ should be
re-defined to ensure convergence. Here we describe the corresponding
procedure, following, in particular, Refs. \cite{Rubtsov_ising,Sun_Kotliar_EDMFT}.

For illustration purposes, we suppose $\alpha$ to be just a real constant
factor; the generalization to a complex matrix $\alpha$ is straightforward.
Let us introduce a basis where $W^{-1}$ is diagonal. Let us number the
states in this basis with an index $l$, and consider some matrix element $%
W_{ll} $. For a positive real part of $W_{ll}$, the corresponding part of the
action can be decoupled as follows:
\begin{equation}  \label{HSplus}
e^{\rho_l^{\prime}W_{ll} \rho_l^{\prime}+\rho_l^{\prime\prime}W_{ll}
\rho_l^{\prime\prime}}=\alpha W_{ll}^{-1} \alpha \int d\eta^{\prime}\int
d\eta^{\prime\prime}e^{\eta^{\prime}_l \alpha
\rho^{\prime}_l-\eta^{\prime\prime}_l \alpha
\rho^{\prime\prime}_l-\eta^{\prime}_l \alpha W_{ll}^{-1} \alpha
\eta^{\prime}_l-\eta^{\prime\prime}_l \alpha W_{ll}^{-1} \alpha
\eta^{\prime\prime}_l},
\end{equation}
where we explicitly introduced the Hermitian and anti-Hermitian parts of $%
\rho=\rho^{\prime}+i \rho^{\prime\prime}$. One can see that for $\eta_l^{\prime}$ and
$\eta_l^{\prime\prime}$ being interpreted as real and
imaginary parts of a complex number $\eta$, this decoupling becomes an
explicit form of the second line of Eq. (\ref{HS}), written for a particular
state $l$. The integral limits for positive $W_{ll}$ can be chosen along the
real axis, from $-\infty$ to $+\infty$ for both $\eta_l^{\prime}$ and $%
\eta_l^{\prime\prime}$.

The key observation is that for $W_{ll}$ having a \textit{negative} real
part, the decoupling (\ref{HSplus}) still holds, up to the integration path
along imaginary axis, from $-i \infty$ to $+i \infty$ for $\eta_l^{\prime}$
and in the opposite direction, that is $+i \infty$ to $-i \infty$, for $%
\eta_l^{\prime\prime}$.
Thus, the integration over $%
D[\eta^*, \eta]$ can be considered as a symbolic notation for the
integration over $D[\eta^{\prime}, \eta^{\prime\prime}]$ with $%
\eta=\eta^{\prime}+i \eta^{\prime\prime}, ~ \eta^*=\eta^{\prime}-i
\eta^{\prime\prime}$ substituted in the integrand. The integration path over
each component of vectors $\eta^{\prime}, \eta^{\prime\prime}$ is explicitly
defined for the basis where $\alpha W \alpha$ is diagonal: the integration
goes along the real or imaginary axis, corresponding to the sign of the real part of
the diagonal matrix element  $W_{ll}$.

One can check that such a definition of $\int D[\eta^*, \eta]$ still allows one
to use the standard algebraic manipulations of path integrals. In particular,
to obtain the Green's function of dual bosonic variables $\tilde{G}=-i <\eta \cdot
\eta^*>$ one uses the standsrd procedure of differentiation with respect to
$\alpha W^{-1} \alpha$, as can be explicitly shown from Eq. (\ref{HSplus}).
On the other hand, the statistical properties of dual variables can be
unusual because of the imaginary axis integration path. For instance, similarly to
the fermionic dual Green's function \cite{Rubtsov_big_PRB}, $<\eta \cdot
\eta^*>=<\eta^{\prime}\cdot \eta^{\prime}>+ <\eta^{\prime\prime}\cdot
\eta^{\prime\prime}>$ is not necessary positive-defined and consequently the imaginary part of its Fourier transform
 is not always negative.

Let us continue with the derivation of the dual formalism.
The Hubbard-Stratonovich decoupling in the partition function with the action (%
\ref{Lattice}) gives (in matrix form)
\begin{eqnarray}\label{Smix}
\begin{array}{l}
\int D[c^{\dag }c]e^{-\sum_{r}S_{imp}[c_{r}^{\dag },c_{r}]+c^{\dag }(\Delta
-\varepsilon )c+\rho ^{\ast }(\Lambda -V)\rho }=\det (\alpha _{f}^{-1}(\Delta
-\varepsilon )\alpha _{f}^{-1})\det (\alpha _{b}(\Lambda -V)^{-1}\alpha
_{b})\times \\
\\
\times \int D[c^{\dag }c]\int D[f^{\dag }f]\int D[\eta ^{\ast }\eta
]e^{-\sum_{r}S_{imp}[c_{r}^{\dag },c_{r}]+f^{\dag }\alpha _{f}c+c^{\dag
}\alpha _{f}f+\rho ^{\ast }\alpha _{b}\eta +\eta ^{\ast }\alpha _{b}\rho
-f^{\dag }\alpha _{f}(\Delta -\varepsilon )^{-1}\alpha _{f}f-\eta ^{\ast
}\alpha _{b}(\Lambda -V)^{-1}\alpha _{b}\eta }.
\end{array}
\end{eqnarray}
Integrating out the ``localized'' fermionic Grassmann variables $c^{\dag },c$
with $\alpha =g_{\omega}^{-1} $ for the fermionic and $\alpha =\chi _{\Omega }^{-1}$ for the
bosonic part, we arrive at the following action in dual variables
\begin{equation}
\begin{array}{l}
\tilde S=-\sum_{\omega k}\tilde{\mathcal{G}}_{\omega k}^{-1}f_{\omega K}^{\dag
}f_{\omega K}-\sum_{\Omega k}\tilde{\mathcal{X}}_{\Omega K}^{-1}\eta
_{\Omega K}^{\ast }\eta _{\Omega K}+\sum_{i}\tilde{U}[\eta
_{i},f_{i},f_{i}^{\dag }] \\
\\
\tilde{\mathcal{G}}_{\omega k}=(g_{\omega}^{-1}+\Delta _{\omega }
  -\varepsilon _{k})^{-1}-g_{\omega }\\
\\
\tilde{\mathcal{X}}_{\Omega K}=(\chi _{\Omega}^{-1}+\Lambda _{\Omega }
-V_{K} )^{-1}-\chi _{\Omega } \\
\\
\tilde{U}[\eta ,f,f^{\dag }]=\sum_{\omega \Omega }\left( \lambda _{\omega
\Omega }\eta _{\Omega }^{\ast }f_{\omega +\Omega }^{\dag }f_{\omega
}+\lambda _{\omega \Omega }^{\ast }\eta _{\Omega }f_{\omega }^{\dag }\
f_{\omega +\Omega }\right) +\frac{1}{4}\sum_{\omega \omega ^{\prime }\Omega
}\gamma _{\omega \omega ^{\prime }\Omega }f_{\omega +\Omega }^{\dag
}f_{\omega ^{\prime }-\Omega }^{\dag }f_{\omega  ^{\prime }}f_{\omega}+...,
\end{array}
\label{DB_action}
\end{equation}
where $\gamma$ is a full four-point fermionic vertex of the impurity model
\begin{equation}  \label{Gamma}
\gamma_{\omega \omega^{\prime}\Omega}=\frac{<c_{\omega+\Omega}
c_{\omega^{\prime}-\Omega} c^\dag_\omega
c^\dag_{\omega^{\prime}}>_{imp}-g_\omega g_{\omega^{\prime}}
(\delta_{\Omega+\omega-\omega^{\prime}}-\delta_{\Omega})}{g_{\omega+\Omega}
g_{\omega^{\prime}-\Omega} g_\omega g_{\omega^{\prime}}}
\end{equation}
($\delta$ is the Kronecker symbol), and $\lambda$ is a ``mixed'' quantity
\begin{equation}  \label{Lambda}
\lambda_{\omega \Omega}=\frac{-<c_{\omega+\Omega} c_\omega^\dag
\rho_\Omega>_{imp}-<\rho>_{imp} g_\omega \delta_\Omega} {g_\omega
g_{\omega+\Omega} \chi_\Omega}.
\end{equation}
The second term in the nominator typically equals zero as $<\rho>_{imp}$
vanishes.
The values of $\gamma$ and $\lambda$ are related, since $\rho_\Omega=\sum_{%
\omega^{\prime}}  c^\dag_{\omega^{\prime}\sigma} s_{\sigma \sigma^{\prime}}
c_{\omega^{\prime}-\Omega, \sigma^{\prime}}$,
\begin{equation}  \label{lambdagamma}
\lambda_{\Omega \omega}=\chi_\Omega^{-1}\left(1-\sum_{\omega^{\prime} \sigma \sigma^{\prime}}
\gamma^{\sigma \sigma^{\prime}}_{\omega \omega^{\prime} \Omega} g^{\sigma}_{\omega^{\prime}}
s_{\sigma \sigma^{\prime}} g^{\sigma^{\prime}}_{\omega^{\prime}-\Omega} \right)
\end{equation}
It is important to note that for the Gaussian impurity problem $\gamma$
vanishes, but $\lambda$ takes a finite value of $\chi_\Omega^{-1}$.

The following exact relations between fermionic and bosonic Green's functions of
original and dual variables can be proven:
\begin{eqnarray}
 \label{Exact}
 \begin{array}{l}
G_{\omega k}=(\Delta _{\omega }-\varepsilon _{k})^{-1}g_{\omega }^{-1}\tilde{G}%
_{\omega k}g_{\omega }^{-1}(\Delta _{\omega }-\varepsilon _{k})^{-1}+(\Delta
_{\omega }-\varepsilon _{k})^{-1}; \\
\\
X_{\Omega K}=(\Lambda _{\Omega }-V_{K})^{-1}\chi _{\Omega }^{-1}\tilde{X}%
_{\Omega K}\chi _{\Omega }^{-1}(\Lambda _{\Omega }-V_{K})^{-1}+(\Lambda
_{\Omega }-V_{K})^{-1};
\end{array}
\end{eqnarray}
The relation for the fermionic Green's functions is derived in Ref. \cite
{Rubtsov_big_PRB}. Main idea is to substitute the right-hand side of Eq.(\ref
{Smix}) in the formula $G_{\omega k}=-\frac{1}{Z}\frac{\partial Z}{\partial
\varepsilon _{\omega k}}$. The last term $(\Delta _{\omega }-\varepsilon
_{k})^{-1} $ appears from the differentiation of the fermionic determinant. The
bosonic relation is derived quite similarly, but the signs appearing during
the derivation should be discussed. First, the expression for the bosonic
Green's function has a different sign, $X_{\Omega k}=\frac{1}{Z}\frac{%
\partial Z}{\partial V_{\Omega k}}$. This gives factors of  $-1$ at $X$ and $\tilde{X}$.
Second, since the bosonic Hubbard-Stratonovich transformation
contains an inverse of the determinant, the same different sign appears for
the last term $(\Lambda _{\Omega }-V_{k})^{-1}$, so that the signs of all
the terms are changed, compared to the fermionic relation between the real and
dual fermionic Green's functions. Therefore the entire resulting relationship
between the real and dual bosonic Green's functions obeys exactly the same
form, similar to the results of the supersymmetric Hubbard-Stratonovich
transformation \cite{Stanescu_Kotliar} .

It is useful to introduce the self-energy (for fermions) and
polarization-operator (for bosons) corrections to the effective-medium
approximation, defined as
\begin{equation}  \label{SelfEnergy}
\Sigma^{\prime}=G^{-1}-\mathcal{G}^{-1}; ~~~ \Pi^{\prime}=X^{-1}-\mathcal{X}%
^{-1}.
\end{equation}
We remind that $\mathcal{G}, {\mathcal{X}}$ are given by the expressions (%
\ref{MFG}).

The self-energy and polarization operator of the dual ensemble are given by the
following expressions:
\begin{equation}  \label{SelfEnergyDual}
\tilde{\Sigma}=\tilde{G}^{-1}-\tilde{\mathcal{G}}^{-1};~~~ \tilde{\Pi}=%
\tilde{X}^{-1}-\tilde{\mathcal{X}}^{-1}.
\end{equation}
Being re-expressed in these quantities, the exact relations (\ref{Exact})
have a particularly simple form:
\begin{eqnarray}  \label{Exact2}
\begin{array}{l}
(\Sigma^{\prime}_{\omega k})^{-1}=\tilde{\Sigma}_{\omega k}^{-1}+g_\omega
\\
\\
(\Pi^{\prime}_{\Omega K})^{-1}=\tilde{\Pi}_{\Omega K}^{-1}+\chi_\Omega
\end{array}
\end{eqnarray}
which results in the following expression for the fermionic and bosonic Green's functions:
\begin{eqnarray}
 \label{Exact}
 \begin{array}{l}
G_{\omega k}^{-1}=
(g_{\omega}+g_{\omega} \tilde{\Sigma}_{\omega k}  g_{\omega} )^{-1}+\Delta _{\omega }
  -t_{k}
\\
\\
X_{\Omega K}^{-1}=
(\chi _{\Omega}+\chi _{\Omega}  \tilde{\Pi}_{\Omega K}     \chi _{\Omega}     )^{-1}+\Lambda _{\Omega }
-V_{K} \\
\end{array}
\end{eqnarray}
One can immediately see from these formulas that the effective-medium
expressions (\ref{MFG}) correspond to the Gaussian approximation for the
dual variables: $\tilde{\Sigma}, \tilde{\Pi}$ (and consequently $\Sigma^{\prime}, \Pi^{\prime}$) vanish.

Following the effective medium paradigm we introduce a generalized self-consistent
condition for
$\Delta _{\omega }$ and $\Lambda _{\Omega }$ similar to the Eq.(\ref{DMFTcondition}), but
for "exact" fermionic and bosonic Green's functions:
\begin{eqnarray}\label{DBcondition}
g_\omega=\sum_k { G}_{\omega k}, \\ \nonumber
\chi_\Omega=\sum_K { X}_{\Omega K}
\end{eqnarray}

\section{Basic diagrams and simple approximations}
\label{BareDiagSec}

The simplest non-local physics can already  be described in the EDMFT.
In our formalism this corresponds to a zero-th order approximation
(``no diagrams'') $ \tilde{\Sigma}=0$ and $\tilde{\Pi}=0$, and
condition (\ref{DMFTcondition}) is satisfied. We intend to preserve the
advantages of the EDMFT, but also include  the formation of collective modes and their
interaction with the electrons. For a ``purely fermionic''
Hubbard-like theory this means an account of the ladder series in dual
space \cite{Hafermann_Ladder}.

\begin{figure}[tbp]
\begin{center}
\includegraphics[width=30mm]{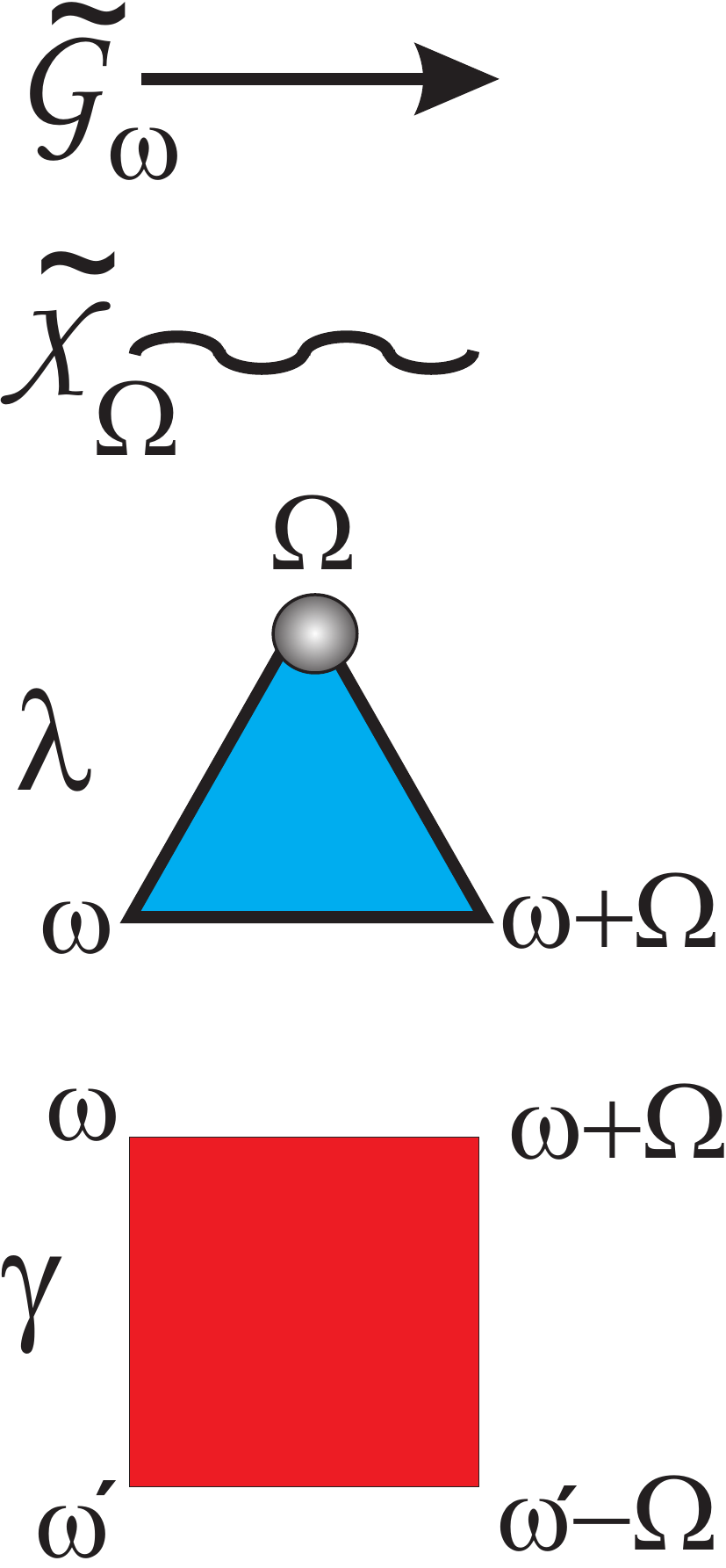}
\end{center}
\caption{Basic building blocks for dual-boson diagram:
fermionic ($\tilde{\mathcal{G}}$) and bosonic ($\tilde{\mathcal{X}}$)
dual propagators as well as $\lambda$ and $\gamma$ vertices.}
\label{Basic_diagram}
\end{figure}

\begin{figure}[tbp]
\begin{center}
\includegraphics[width=100mm]{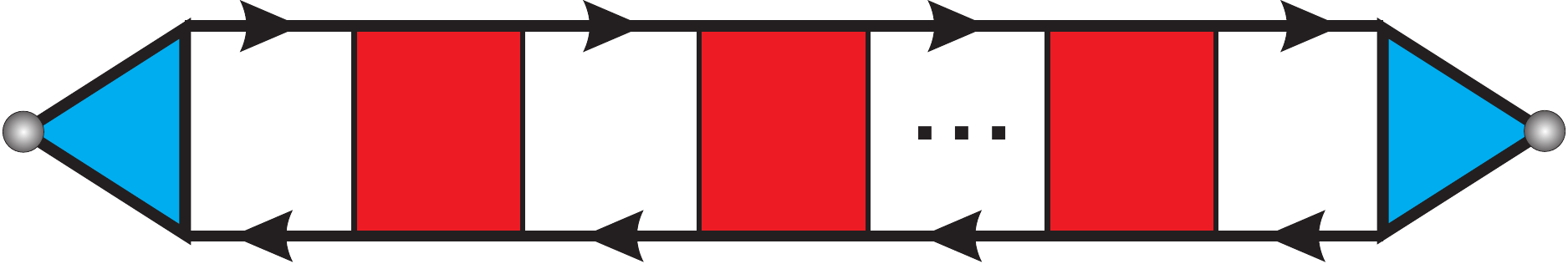}
\end{center}
\caption{ The bosonic dual self-energy in the ladder
approximation. A triangle represents the $\protect\lambda$ vertex and a square
represents the $\protect\gamma$ vertex}
\label{fig_ladder}
\end{figure}

\begin{figure}[tbp]
\begin{center}
\includegraphics[width=50mm]{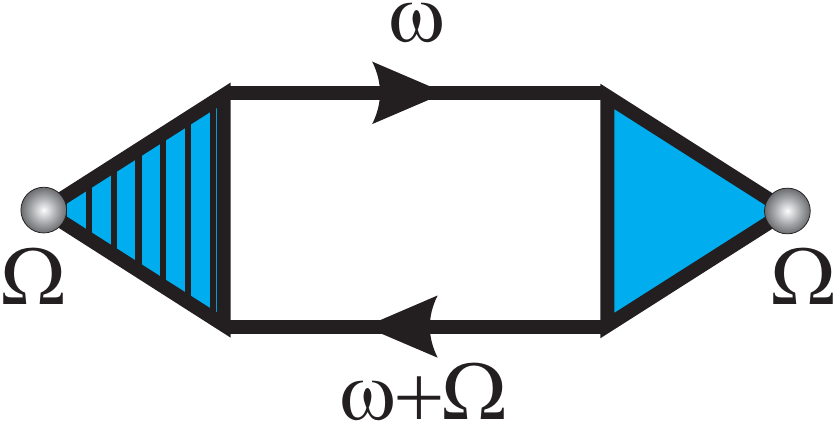}
\end{center}
\caption{ The bosonic dual self-energy with the renormalized triangle vertex.}
\label{fig_pi}
\end{figure}

For the present theory with the dual action of Eq.(\ref{DB_action}),
there are two basic Green's function lines (Fig. \ref{Basic_diagram}):
a fermionic (full-line with arrow) and a bosonic (wavy-line) as well
as two interaction vertices: a triangle (corresponding to the dual
boson-fermion interaction - $\lambda $) and a square (describing
the dual-fermion interaction - $\gamma $). In this paper, we restrict ourselves to a polarization operator, given by
the electron ladder ended by two triangles (Fig. \ref{fig_ladder} ). Equally, one can consider this diagram as
 just a two-lines bubble, with a renormalized vertex (Fig. \ref{fig_pi} ).
\begin{equation}  \label{buble}
\begin{array}{l}
\tilde \Pi_{\Omega K}= { \lambda}^{\rm{eff}}_{\Omega} \cdot \tilde{{\tt X}}_{\Omega K}^0 \cdot \lambda_{\Omega},\\   \\
\lambda^{\rm{eff}}_\Omega=\left(1-\gamma_\Omega \tilde{{\tt X}}_{\Omega K}^0 \right)^{-1} \lambda_\Omega .
\end{array}
\end{equation}
Here a symbolic notation is used, so that $\gamma_\Omega$ is a tensor, and $\lambda_\Omega$ is a vector with components corresponding to different $\omega$, and the dots in the first line
denote a scalar product vector-tensor-vector. The bare dual polarization operator (an empty loop of the dual Green's function)
 $\tilde {\tt X}^{0}$ is a tensor having only diagonal components. Later we also use bare polarization operators for the initial variables and for the impurity model, so that
\begin{equation}
\begin{array}{l}
\tilde{{\tt X}}_{\Omega K}^{0,\omega}=-\sum_{k} {\tilde{\cal G}}_{\omega k} {\tilde{\cal G}}_{\omega+\Omega k+K},\\ \\
{{\tt X}}_{\Omega K}^{0,\omega}=-\sum_{k} {{\cal G}}_{\omega k} {{\cal G}}_{\omega+\Omega k+K},\\ \\
{{\tt x}}_{\Omega}^{0,\omega}=-{{g}}_{\omega} {{ g}}_{\omega+\Omega}.
\end{array}
\end{equation}
Scalar polarizabilities, respectively denoted by ${\tilde {\cal X}}^{0}_{\Omega K}$, ${{\cal X}}^{0}_{\Omega K}$, and $\chi^0_\Omega$, can be found by summating over fermionic frequencies, so that, for example,  ${{\cal X}}^{0}_{\Omega K}=-\sum_{k\omega} {{\cal G}}_{\omega k} {{\cal G}}_{\omega+\Omega k+K}$.

The simplest approximation one can make is just to neglect the vertex part $\gamma$.
In this case
$\lambda^{\rm{eff}}_{\omega\Omega}=\lambda_{\omega\Omega}=\chi^{-1}_\Omega$. Simple manipulations
with the second lines of Eqs. (\ref{MFG}), (\ref{SelfEnergy}) and
(\ref{Exact2}) then give
\begin{equation}  \label{DRPA}
X_{\Omega K}=\frac{1}{\left(\chi_\Omega+\tilde{{\cal X}}^{0}_{\Omega K}\right)^{-1} +\Lambda_\Omega-V_K}.
\end{equation}
A similar non-local contribution to the fermionic self-energy can be introduced to
describe the renormalization of the fermionic Green's function $G$.
The corresponding diagram contains one fermionic and one bosonic line (Fig. \ref{fig_sig}).
We will however concentrate on the analysis of bosonic quantities and
for simplicity use the EDMFT result $\tilde{\mathcal{G}}$ instead of
the renormalized $\tilde G$ propagator. This allows for a further simplification.
Indeed, in the local approximation $\mathcal{G}$ and $\tilde{%
\mathcal{G}}^{dual}$ differ in their local part only \cite
{Rubtsov_DF,Rubtsov_big_PRB}:
\begin{equation}  \label{GGprime2}
\tilde{\mathcal{G}}_{\omega k}=\mathcal{G}_{\omega k}-g_{\omega}.
\end{equation}
Further, since we suppose that $\gamma=0$ and since $<\rho>=0$, the bosonic
Green's function for the impurity problem becomes equal $\chi_\Omega^0$. Substituting
the expression (\ref{GGprime2}) into Eq. (\ref{DRPA}), we obtain
\begin{equation}  \label{RPA2}
X_{\Omega K}=\frac{1}{\left({\mathcal X}^{0}_{\Omega K}\right)^{-1}+ \Lambda_\Omega-V_K }.
\end{equation}
For $\Lambda=0$, this expression is clearly related to the Random Phase
Approximation (RPA). To get the ``textbook'' RPA, one should completely
neglect the interaction terms of the impurity problem, so that $%
g_\omega=(i\omega-\Delta_\omega)^{-1}$ and $\mathcal{G}_{\omega,
k}=(i\omega-\varepsilon_k)^{-1}$. The scheme with an exactly calculated
$g_\omega$ can be called RPA+DMFT.
In this case the RPA is a simple case of Fig. \ref{fig_pi}.
To go beyond the RPA, one should be careful. As we will discuss in Sec. \ref{conserv}
in general it is wrong to work with the renormalized Green's functions but to neglect the vertex $\gamma$.
Because this may lead to a violation of the conservation laws.

A theory using the single-bubble approximation (\ref{buble}) with
 $\lambda^{\rm{eff}}$ approximated by $\lambda$ (that is, without putting $\lambda$ to
$\chi^{-1}$) is essentially the GW+EDMFT approach
\cite{Sun_Kotliar_EDMFT}. There is only a slight difference in the
self-consistency condition, as we explain in Appendix. It is also
important that for the diagram calculation of the GW+EDMFT approach that the
local part of the bosonic Green's function is excluded \textit{ad
hoc}. As one can see from Eq. (\ref{EDMFTGGprime}), for the EDMFT
choice of the Bose variables, it actually does not vanish exactly. So,
our consideration proposes a certain a modification of GW+EDMFT.
Anyhow, such modification should not strongly affect the results of the
GW+EDMFT. The real advantage of our consideration is the
possibility to go beyond the domain of applicability of the
GW+EDMFT, i.e. in  is the situation when the single-bubble approximation
becomes incorrect. It is considered in the next sections.

\begin{figure}[tbp]
\begin{center}
\includegraphics[width=75mm]{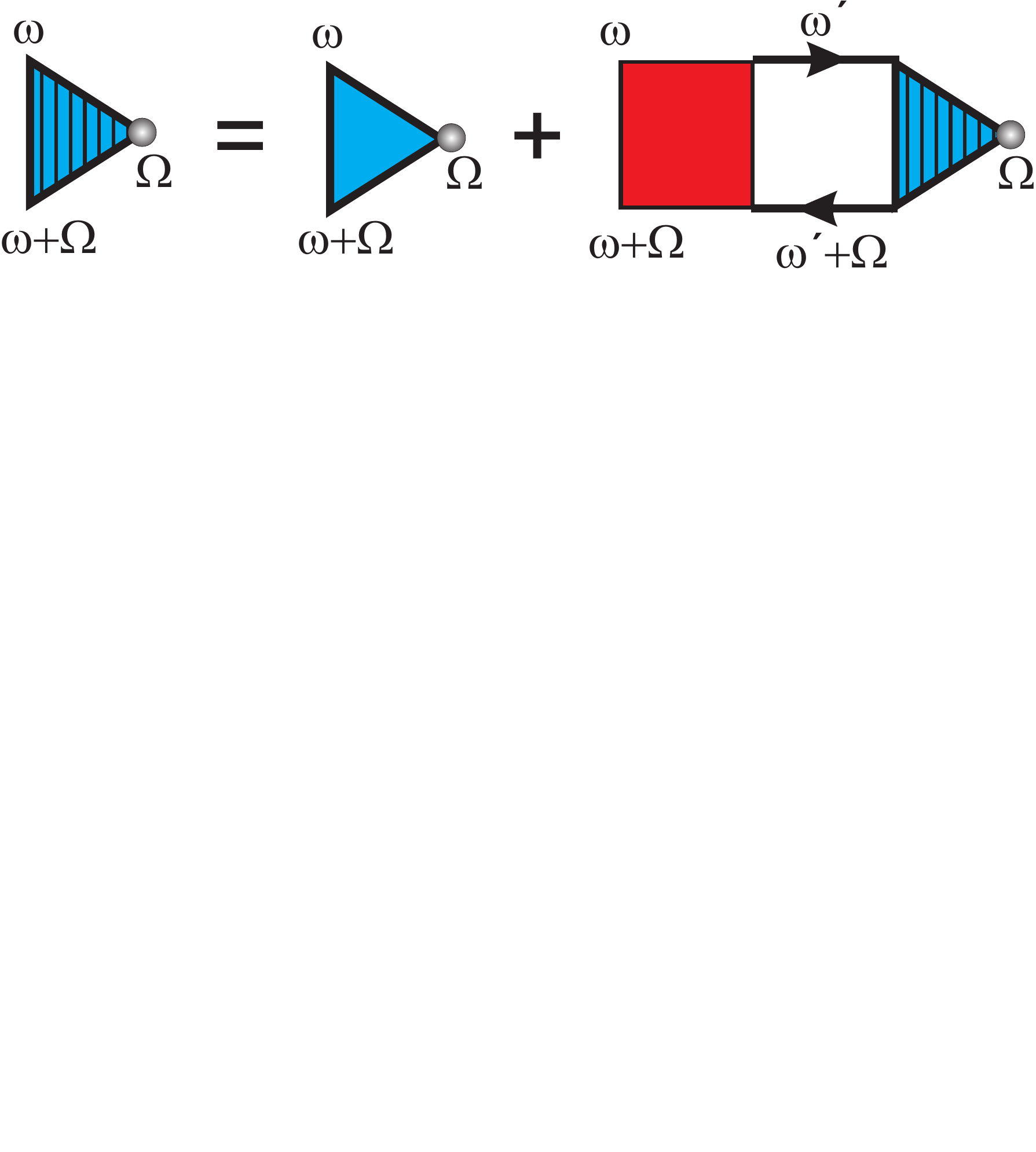}
\end{center}
\caption{ The diagrammatic equation for the renormalized triangle vertex.}
\label{bold3}
\end{figure}

\begin{figure}[tbp]
\begin{center}
\includegraphics[width=50mm]{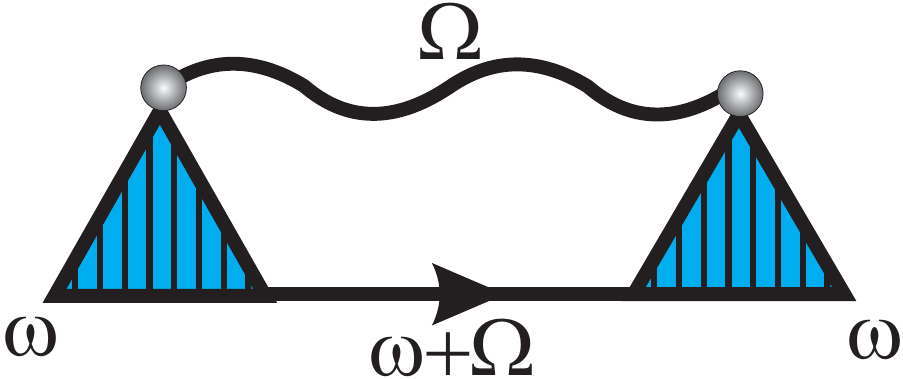}
\end{center}
\caption{ An example of diagram for the fermionic dual self-energy with the renormalized triangle vertex.}
\label{fig_sig}
\end{figure}

The diagrams shown in Figs. \ref{bold3}, \ref{fig_pi}, \ref{fig_sig} form the
minimal consistent set beyond the RPA to describe the formation of collective modes
with electron-boson coupling taken into account explicitly.

\section{Charge conservation}
\label{conserv}

So far, we did not consider whether the constructed approximations conserve particle number,
total spin, etc. However, the conservation laws are particularly important for
bosonic excitations. For example, the charge (particle number) conservation resulting in the equation $\Omega^2 <\rho \rho>_{\Omega K}=K^2 <j j>_{\Omega K} $  immediately leads
to the requirement $ <\rho \rho>_{K=0}=0$ at any finite frequency (provided that the current-current correlator
$<j j>_{\Omega, K=0}$ is finite). For 3D systems with Coulomb interaction the long-wavelength asymptotic
behaviour for the density-density correlator should be
 $ <\rho \rho>_{\omega, K\to 0}\propto \frac{K^2}{\Omega^2+\Omega_p^2}$, where $\Omega_p$ is the plasma frequency. The RPA for free electrons is proven to obey this property, since $X^{0}_{\Omega, K\to0}$ vanishes with ${\cal G}=(i\omega-\varepsilon_k)^{-1}$, but this is not true for the renormalized Green's function \cite{Baym}. Indeed, following the standard proof \cite{AGD, Mahan} one writes  ${{\cal X}}^{0}_{\Omega K}=-\sum_{k\omega} {{\cal G}}_{\omega k} {{\cal G}}_{\omega+\Omega k+K}$ as
\begin{equation}
{\mathcal X}^{0}_{\Omega K}=-\sum_{\omega k} \left( \frac{1}{i\omega-\varepsilon_k-\Sigma_\omega}-
\frac{1}{i(\omega+\Omega)-\varepsilon_{k+K}-\Sigma_{\omega+\Omega}}\right)\frac{1}{i\Omega+\Sigma_\omega-\Sigma_{\omega+\Omega}}.
\end{equation}
For free electrons of dispersionless $\Sigma$ (that is, in static mean-field) the summation
over $\omega$ produces a difference
of occupation numbers, $n_k-n_{k+K}$, in the nominator, that vanishes at   $K=0$. However this
reasoning clearly breaks down for the frequency-dependent $\Sigma_\omega$ in DMFT approximation.
It was known for a long time that the neglection  of vertex corrections together with the use of {\it renormalized} Green's
functions, violate charge and spin conservations \cite{Baym,Baym_hist,Hertz}.
We found that
$X^{0}$ with the DMFT Green's function does not vanish at small $K$  and therefore it follows form Eq. (\ref{RPA2}) that  {\it EDMFT+GW is not a valid description of plasmons}. All the more, {\it plasmons do not appear in simple EDMFT}, as one can easily observe from the second line of Eq. (\ref{MFG}): the local quantity $\chi_\Omega^{-1}+\Lambda_\Omega$ cannot behave like $\Omega^2 K^2$. This is the reason why the GW approach is normally used only for calculating the single-particle Green's function, while any two-particle properties are
investigated within the Bethe-Salpeter equations \cite{GW_RMP}. Nevertheless, such a scheme cannot describe self-consistent effects of collective excitations on the single-particle spectral function.

To construct a conserving approximation, it is useful to consider the following reformulation of the theory. Let us return to the initial lattice action (\ref{Lattice}) and do the Hubbard-Stratonovich decoupling for the bosonic variables only. It will give an analog of  Eq. (\ref{Smix}) with the action
\begin{equation}\label{SFmix}
\begin{array}{l}
S=S^F[c,c^\dag]+\rho^* \alpha_b \eta +\eta^* \alpha_b \rho-\eta^* \alpha_b (\Lambda-V)^{-1} \alpha_b \eta,\\
S^F=\sum_r S_{at}[c^\dag_r, c_r]+\sum_{r,R\neq 0, \omega, \sigma}
\varepsilon_R c^\dag_{r \omega \sigma} c_{r+R \omega \sigma} +\sum_{r, \Omega} \Lambda_{\Omega} \rho_{r \Omega}^*\rho_{r \Omega}.
\end{array}
\end{equation}
Physically, $S^F$ describes a lattice with the interaction being local in space but retarded. Now we formally integrate over the fermionic variables. The Gaussian approximation in dual bosons for such a theory corresponds to the expression
\begin{equation}\label{GreenF}
{X}_{\Omega K}=\frac{1}{\left(X^F_{\Omega K}\right)^{-1}+\Lambda_\Omega-V_K},
\end{equation}
where $X^F$ is an exact $K$-dependent two-particle correlator for the lattice problem $S^F$.
Formally, the calculation of $X^F_{K, \Omega}$ corresponds to a summation of all parts of the diagrams
with fermionic lines in series for Eq.(\ref{DB_action}). Comparing Eqs. (\ref{GreenF}) and  (\ref{Exact}) we obtain
\begin{equation}\label{XPi}
\tilde{\Pi}_{\Omega K}=\chi^{-1}_\Omega X^F_{\Omega K} \chi^{-1}_\Omega-\chi^{-1}_\Omega.
\end{equation}

This way, we need a conserving approximation for $X^F$. Such an approximation is actually known within the framework of DMFT. Because it is a conserving theory in the Baym-Kadanoff sense \cite{Baym}, with the self energy being just the sum of all local diagrams. In such a theory, the conserving approximation for two-particle quantities can be obtained as a Green's function variational derivative with respect to the dispersion law,
\begin{equation}\label{susc4}
<(c_1 c_{1'}^\dag+g_{11'})(c_2 c_{2'}^\dag+g_{22'})>=\frac{\delta G_{11'}}{\delta \varepsilon_{22'}}.
\end{equation}
An important property of the Baym's  criterion of consistency is that the variation should be considered self-consistently, so that the $\Delta$ and $g$ of DMFT vary with $\varepsilon$ to preserve the self-consistency condition (\ref{DMFTcondition}).
One can also explicitly see that such a self-consistent variation preserves the charge conservation. Indeed, the charge conservation follows from gauge invariance under the transformation $c_{\tau, r}\to c_{\tau, r} e^{-i \delta \phi_{\tau,r}},~~c_{\tau, r}^{\dag}\to c_{\tau, r}^{\dag} e^{i \delta \phi_{\tau,r}}$ with an arbitrary dependence of the infinitesimal phase shift $\delta\phi$ on time and space arguments. In an exact theory, such a transformation just results  in similar phase rotations for averages, for example $G_{11'}\to G_{11'} e^{i(\phi'-\phi)}$. A conservative approximation must inherit this property.
DMFT is entirely defined by the self-consistency condition (\ref{DMFTcondition}) and impurity action (\ref{Impurity}). If all the quantities entering the Green's function ${\cal G}_{\omega k}$
(such as $\varepsilon$, $\Delta$, $g$) evolve simultaneously under the gauge transformation,
the self-consistency (\ref{DMFTcondition}) clearly stays fulfilled. The entire Green's function ${\cal G}_{\omega k}$ obeys the gauge phase rotation, which proves that the theory is conserving.

We can prove that the ladder approximation for dual bosons results in such a conserving theory. We introduce a variation of
the local part of the dispersion law
$\delta \varepsilon_{t t' r r}= e^{i (K r-\Omega t)} \delta \varepsilon_{\Omega K} (t-t')$, and corresponding variations of local quantities  $\delta \Delta_{t t' r}= e^{i (K r-\Omega t)} \delta \Delta_{\Omega K} (t-t')$, and $\delta g_{t t' r}= e^{i (K r-\Omega t)} \delta g_{\Omega K} (t-t')$. In the tensorial notation used  previously, the variations $\delta \varepsilon_{\Omega K}, \delta \Delta_{\Omega K}, \delta g_{\Omega K}$ are vectors with a single index  $\omega $ ; their components are  defined as Fourier-transforms: $\delta \varepsilon_{\Omega K}^\omega=\int \delta \varepsilon_{\Omega K}(\tau) e^{i\omega\tau} d\tau$ etc.

The straightforward variation of the self-consistency condition with the DMFT Green's function (\ref{DMFTcondition}) gives:
\begin{equation} \label{suscdraft}
\delta {g}_{\Omega K}={\tt X}^{0}_{\Omega K}
\left(\frac{\delta g_{\Omega K}}{{\tt x}^{0}_{\Omega}} +\delta\Delta_{\Omega K}-\delta\varepsilon_{\Omega K} \right).
\end{equation}
It is worth to remind that ${\tt X}^0_{\Omega K}$ and ${\tt x}^0_\Omega$ are tensors with double indices $(\omega, \omega' )$, as well as the quantities  ${\tt X}^F_{\Omega K}$ and ${\tt x}_\Omega$ to appear in the following formulas.

Now, let us consider the full lattice susceptibility ${\tt X}^F_{\Omega K}$ as a Fourier-transform of Eq.(\ref{susc4}) with pairwise-equal space arguments:
\begin{equation}
{\tt X}^{\omega_1 \omega_2}_{\Omega K}=\sum_{r_1,r_2}\int{\frac{\delta G_{t_1,t_1',r_1,r_1}}{\delta \varepsilon_{t_2,t_2',r_2,r_2}}} e^{i\left(\omega_1 (t_1-t_1')-\omega_2(t_2-t_2') +\Omega (t_1-t_2)-K (r_1-r_2)\right)}
dt_1dt_1'dt_2dt_2'.
\end{equation}
The self-consistency requires $\delta G_{t_1,t_1',r_1,r_1}=\delta g_{t_1,t_1',r_1}$, and therefore the above-introduced variations correspond to
\begin{equation}
{\tt X}^{F}_{\Omega K}=-\frac{\delta {g}_{\Omega,K}}{\delta \varepsilon_{\Omega,K}}.
\end{equation}
Similarly, for the susceptibility of the impurity model
\begin{equation}
{\tt x}_{\Omega}=-\frac{\delta g_{\Omega K}}{\delta \Delta_{\Omega K}}.
\end{equation}
Although the right-hand side formally includes the wavevector $K$, $\tt x$ is of course $K$-independent.

With the above two formulas formulas, it immediately follows from (\ref{suscdraft}) that
\begin{equation}\label{susc}
\frac{1}{{\tt X}^F_{\Omega K}}-\frac{1}{{\tt x}_{\Omega}}=
\frac{1}{{\tt X}_{\Omega K}^{0}}-\frac{1}{{\tt x}_{\Omega}^{0}}.
\end{equation}
Such a self-consistent condition (\ref{susc}) is equivalent to locality of the irreducibly vertex and it is used in the DMFT scheme for calculations of non-local susceptibilities \cite{Georges_RMP}. It is worthwhile to remind that the susceptibilities with a single frequency argument, which we defined previously, can be obtained by summation over the fermionic frequencies, so that
$X_{\Omega K}^F=\sum_{\omega_1, \omega_2} {\tt X}^{\omega_1 \omega_2}_{\Omega K}$ and
$\chi_{\Omega }=\sum_{\omega_1, \omega_2} {\tt x}^{\omega_1 \omega_2}_{\Omega }$

To establish a connection with the dual formalism, we substitute this expression into Eq. (\ref{XPi}) and
express ${\tt X}$ using Eq. (\ref{susc}). Simple algebraic manipulations give
\begin{equation}
\tilde {\Pi}_{\Omega K}=\sum_{\omega_1\omega_2} \left(
\frac{{\tt x}_\Omega}{\chi_\Omega} \frac{1}{{\tt x}_\Omega^{0}}
\frac{ {\tt X}^{0}_{\Omega K}-{\tt x}^{0}_{\Omega}}
{1-\frac{1}{{\tt x}_\Omega^{0}} ({\tt x}_\Omega-{\tt x}^{0}_{\Omega}) \frac{1}{{\tt x}_\Omega^{0}} \left({\tt X}^{0}_{\Omega K}-{\tt x}^{0}_{\Omega}\right)}
\frac{1}{{\tt x}_\Omega^{0}} \frac{{\tt x}_\Omega}{\chi_\Omega}
\right)_{\omega_1 \omega_2}.
\end{equation}
One can recognize that $\frac{{\tt x}_\Omega}{\chi_\Omega} \frac{1}{{\tt x}_\Omega^{0}}$  and $\frac{1}{{\tt x}_\Omega^{0}} ({\tt x}_\Omega-{\tt x}^{0}_{\Omega}) \frac{1}{{\tt x}_\Omega^{0}}$ are, respectively, the
previously introduced quantities $\lambda$ and $\gamma$ with proper indices. Furthermore, it follows from the self-consistency condition (\ref{DMFTcondition}) that
\begin{equation}
{\tt X}^{0}_{\Omega K}-{\tt x}^{0}_{\Omega}=\tilde{{\tt X}}_{\Omega K}^{0}.
\end{equation}
So we conclude that the obtained expression for $\Pi$ is exactly the result of the dual ladder summation (\ref{buble}).

Finally, let us consider the plasmon dispersion given by the equation
\begin{equation}
\chi_\Omega \tilde{\Pi}_{\Omega K} \chi_\Omega+\chi_\Omega=\frac{1}{V_K-\Lambda_\Omega}.
\end{equation}
In the long-wavelength limit, one considers an expansion in $K^2$, in particular $\tilde{\tt X}^{0}_{\Omega K}=\tilde{\tt X}^{(0,0)}_{\Omega}+\tilde{\tt X}^{(0,2)}_{\Omega} K^2+...$. The zeroth-order term in the left-hand side vanishes due to charge conservation. The second-order term results in the dispersion equation for plasmons in a 3D correlated lattice:
 \begin{equation}
\chi_\Omega \left(\lambda_\Omega^{\rm {eff}} \cdot
\tilde{\tt X}^{(0,2)}_{\Omega}  \cdot
\lambda_\Omega^{\rm {eff}}\right) \chi_\Omega =\frac{1}{4\pi e^2}.
\end{equation}
For the 2D case, the right-hand side equals $\frac{K}{2 \pi e^2}$.

While deriving the last equation, we supposed that $\Lambda_\Omega$ stays finite and can be therefore neglected in comparison with $V_{K}$ at small $K$. Note however that this quantity still enters $S^F$ and therefore affects the left-hand side of the dispersion relation, so that the physics of local plasmonic fluctuations is taken into account.

\section{Ladder summation in a strong-coupling limit}

The single-bubble approximation becomes invalid when vertex
$\gamma$ is large. There is a general situation of this kind:
the four-point vertex of an impurity problem with \textit{magnetic
moment formed} appears to have a part displaying a slow dynamics that
give a large contribution at low frequencies. This simply means
that for isolated atoms the magnetic moments are free and, thus,
for {\it almost} isolated atoms (when the energy of interatomic
effective interactions is much smaller than the intraatomic one) they can
be considered as {\it almost} integrals of motion.

To illustrate this statement, let us consider the atomic limit of
the single-band Hubbard model. The single-electron Green's function of the
Hubbard atom at half-filling
\begin{equation}  \label{gatomic}
g_{at}=\frac{-i \omega}{\omega^2+(U/2)^2}
\end{equation}
has a time-scale $\tau_U\propto U^{-1}$, which we call fast.
Contrary, the spin-spin correlator $<s_\tau s_{\tau^{\prime}}>$ is
independent, as one can easily check, of time arguments. So, a
time-scale for $<s_\tau s_{\tau^{\prime}}>$ is the inverse
temperature $\beta$. It means that, apart
from a small time domain $|\tau-\tau^{\prime}|\approx U^{-1}$, the value of $%
<s_\tau s_{\tau^{\prime}}>$ is determined by its non-Gaussian part, which
does not fall as $|\tau-\tau^{\prime}|$ increases. The much slower dynamics of $%
\gamma$ in comparison with $g$ just reflects the fact that a
rotation of the spin does not change the energy of the atom,
whereas the change of the particle number does. Direct calculation
of $\gamma$ for the Hubbard atom does support this observation.
The finite-temperature expressions \cite {Hafermann_Superpert} for
$\gamma_{1234}$ contain a ``singular'' term proportional to $\beta
U^2 \delta_{\Omega0}$. So, the first two time arguments
of $\gamma$ in time-domain must coincide, as well as the last two: $%
\gamma=\gamma_{\tau\tau\tau^{\prime}\tau^{\prime}}$, but the difference $%
\tau-\tau^{\prime}$ can be large.

The properties of the impurity problem are different from those of the isolated
atom. However, for the strong-coupling limit the hybridization is
small. In this case the fast dynamics  roughly stay unchanged, so
that the atomic Green's function (\ref{gatomic}) can be used, and the four-point vertex obeys the form $%
\gamma_{\tau\tau\tau^{\prime}\tau^{\prime}}$. The hybridization can introduce
its own slow timescale $\tau_\Lambda$, so that $\gamma$ does depend only on $%
\tau-\tau^{\prime}$. In the frequency representation, this means
\begin{equation}  \label{gammaSlow}
\gamma=\gamma_{\Omega},
\end{equation}
and at small frequencies $\gamma_\Omega\propto \tau_\Lambda U^2$. Since $%
\gamma$ includes a large factor, we can keep only the irreducible term in the
expressions for $\lambda$ and $\chi$. Note that this is completely opposite
to the case considered in Section \ref{BareDiagSec}.

The terms proportional to $g g$ in Eq. (\ref{Gamma}) are small in $1/U$ and therefore can be neglected.
Then we obtain  the following relation from Eq. (\ref{Gamma}):
\begin{equation}  \label{irreducc}
\chi_\Omega=\chi^{0}_\Omega \gamma_\Omega \chi^{0}_\Omega.
\end{equation}
In the above equation we introduced a ``bare susceptibility'' of the
impurity problem
\begin{equation}\label{chi0}
\chi^{0}_\Omega\equiv -\sum_{\omega \sigma} g_{\omega} g_{\Omega+\omega}.
\end{equation}
Next, we use the relation (\ref{lambdagamma}) and keep only the large factor $\gamma$:
\begin{equation}  \label{irreduc}
\lambda_\Omega=(\chi^{0}_\Omega)^{-1}.
\end{equation}
Note that in the present approximation $\lambda$, like $\gamma$,  only has
single frequency argument.

Let us consider the half-filled Hubbard lattice with the nearest-neighbour
hopping $t$, at large $U$. The low-frequency behaviour of this system is
essentially the  dynamics of the Heisenberg model. The effective-medium
description of spin waves in this case should result in the expression
\begin{equation}  \label{Heizenberg}
\chi_\Omega=\sum_K \frac{1}{\chi^{-1}_\Omega+\Lambda_\Omega-J_K},
\end{equation}
with $J_{ij}=\frac{t^2}{U}$ for nearest neighbors.
Although  this expression is rather simple (in our notation, it is just
$\Pi^{\prime}=J_k$),  it was not reproduced by
any  DMFT-like approximations so far.
The superexchange interaction containing $t^2$ can easily be obtained from a
single loop of the fermionic lines, but the RPA and GW+EDMFT
denominators contain an inverse of this quantity (see
Section \ref{BareDiagSec}), so the structure of the formulas is drastically different.

In our formalism, the desired expression appears from the ladder summation,
that is the calculation for the Eq. (\ref{buble}) and corresponding diagram shown in Fig. (\ref{fig_ladder}). In this
summation, we take into account the ``slow'' vertex (\ref{gammaSlow}) only.
Since $\gamma$ depends only on a single time argument, the calculation is
quite simple. With the exact relation (\ref{Exact2}), we obtain
\begin{equation}
\Pi^{\prime}_{\Omega K}=\left[ \left( \lambda \frac{\tilde{X}^{0}_{\Omega
K}}{1- \gamma_\Omega \tilde{X}^{0}_{\Omega K} }\lambda\right)^{-1}+\chi_%
\Omega\right]^{-1} .
\end{equation}
Finally, we substitute the formulas (\ref{irreducc}), (\ref{irreduc}) for $\chi$ and $\lambda$. This
results in the cancellation of $\gamma$ out of the expressions:
\begin{equation}
\Pi^{\prime}_{\Omega K}= \lambda_\Omega {\tilde{X}^{0}_{\Omega K}} \lambda_\Omega ,
\end{equation}
where $\lambda_\Omega$ from Eq. (\ref{irreduc}) has a meaning of an effective local Stoner
parameter. Both $\chi^{0}_\Omega$ and $\tilde{X}^{0}_\Omega$ obey the
``fast'' dynamics only. To determine the low-energy properties, in which we are
interested, it is enough to calculate it at $\Omega=0$, and to
replace the summation over frequency by an integration. A straightforward calculation of the expression (\ref{chi0}) with
$g=g_{at}$ gives $\chi^{0}_{\Omega=0}=\frac{1}{2 U}$. To calculate
$\tilde{X}^0_{\Omega=0}$, we switch to real-space and use the
strong-coupling limit. It gives the Green's function
$\tilde{\mathcal{G}}= \left(g^{0}\right)^{2} t$ for the nearest neighbors
and $\tilde{\mathcal{G}}=0$ elsewhere. Now, the integration gives
$\tilde{X}_{\Omega=0}=\frac{t^2}{4 U^3}$ for the nearest-neighbors and we
end up with the desired expression (\ref{Heizenberg}). This
derivation can be considered as a generalization of the effective
exchange formula for the magnetically ordered case \cite{Licht_exch,Kats_Licht04}
to the paramagnetic state.

The standard theory of magnons in itinerant-electron systems \cite{Moriya} is based
on the RPA which is supposed to be exact at zero temperature; at finite temperatures,
magnon-magnon and electron-magnon interactions should be taken into account \cite{Moriya,Dzyalos_Kondrat,ourRMP}.
Electron-hole excitations are primary objects in such an approach, spin waves are
coherent superposition of such excitations and, in general, are only well-defined  in
a restricted part of the Brillouin zone, due to Stoner damping. In the case of strong-coupling
Hubbard model, we have an ``anti-RPA'' situation when local magnetic moments are
primary objects. Therefore, completely different approaches should be used, e.g. those based on the
Hubbard X-operators \cite{vons_kats_tref1993b,irir}. It is very non trivial that the dual boson approach
describes  these two opposite cases within the same formalism, thus providing a practical way to interpolate
between weak and strong coupling limits.

\section{Conclusions}

In summary, the general scheme for non-local correlation effects
within the dual boson approach is presented. This method gives a
direct means to obtain the higher-order diagrammatic contributions
beyond the EDMFT scheme free from the double-counting contributions
and it can be useful to describe
the magnon-spectrum renormalization in strongly correlated systems.
The method combines the numerically exact solution of an effective
single-cite impurity model with frequency-dependent
electron-electron interactions as well as an analytical diagrammatic
expansion of non-local fermionic and bosonic self-energies. This
approach is designed to describe of correlated systems with
strong local fermion correlations and fluctuating non-local
bosonic modes on equal footing.

In this paper we focused on the applications to the description of bosonic
degrees of freedom. Additional contributions to the electron self energy also arise,
in comparison with the dual fermion approach \cite{Rubtsov_DF}, e.g., the diagram shown in
Fig. \ref{fig_sig}. Similar diagrams are important in the cases of coexistence of disorder
with electron-electron \cite{altar,altar1} or electron-phonon \cite{anokhin} interactions.
It is important to calculate these additional contributions within
the Hubbard model and make a detailed comparison with dual fermion theory. This
will be the subject of our further studies.

\section*{Acknowledgments}

This work was supported by DFG Grant No. 436 113/938/0-R and RFFI 11-02-01443.
MIK acknowledges a support from the EU-India FP-7 collaboration under MONAMI.
AIL acknowledges a support from SFB 668 and FOR-1346.

\appendix
\section{Different choices of dual variables}

The formalism presented here has an important difference compare to the common
introduction of the EDMFT: whereas we introduce dual bosons $\eta$ by the
Hubbard-Stratonovich decoupling of the quadratic form $(V-\Lambda) \rho^* \rho$,
the term decoupling $V \rho^* \rho$ is decoupled in the standard EDMFT
scheme \cite{Sun_Kotliar_EDMFT,Florens}. As one can see, this does not
result in any physical consequences, as the two approaches yield the identical
results for the averages $\mathcal{G}, \mathcal{X}$ over original variables.
However, properties of the \textit{dual} bosons, of course, do depend on
their definition. In particular, the difference is about local properties of the
dual propagator: for the present formalism, its local part vanishes in the
EDMFT approximation,
\begin{equation}  \label{dual0}
\sum_k \tilde{\mathcal{X}}=0,
\end{equation}
whereas the ``effective-medium'' condition
\begin{equation}  \label{dualEM}
\sum_k \tilde{\mathcal{X}}^V=\chi_\phi
\end{equation}
holds for the decoupling (\ref{Vdecoupl}). Here $\tilde{\mathcal{X}}^V$ is
the Green's function of the dual variables $\phi$ arising from the
decoupling of $V \rho^* \rho$:
\begin{equation}  \label{Vdecoupl}
V \rho^* \rho\to \alpha (\rho^* \phi + \phi^* \rho) - V^{-1} \alpha^2 \phi^*
\phi ,
\end{equation}
and $\chi_\phi\equiv <\phi \phi^*>_{imp}$ is defined for the impurity
problem with decoupled bosonic hybridization, so that
\begin{equation}  \label{ImpDecoupled}
S_{imp}=S_{at}+\Delta_\omega c^\dag_\omega c_\omega + \alpha (\rho^*_\Omega
\phi_\Omega+\phi^*_\Omega \rho_\Omega) - \Lambda^{-1}_\Omega \alpha^2
\phi^*_\Omega \phi_\Omega .
\end{equation}
To maintain a similarity with our formalism, we introduced in the above
expressions a scaling factor $\alpha$, that is absent in the standard EDMFT
scheme \cite{Sun_Kotliar_EDMFT}. All physical quantities do not depend on a
particular value of this factor.

Averages $\tilde{\mathcal{X}}^V$ and $\chi_\phi$ can be found from
the exact relations similar to (\ref{Exact}):
\begin{eqnarray}  \label{ExactV}
\begin{array}{l}
\alpha^2 \tilde{\mathcal{X}}^ V=V \mathcal{X} V+V; \\
\\
\alpha^2 \chi_\phi=\Lambda \chi \Lambda +\Lambda.
\end{array}
\end{eqnarray}

We will now consider a general case of some approximation utilizing a
decoupling of $W \rho^* \rho$ with an arbitrary $W$. The approximation can
be non-Gaussian, so that there is a correction $\Pi^{\prime}$ to the EDMFT
polarization operator:
\begin{equation}  \label{W1}
X=\frac{1}{\chi^{-1}+\Lambda-V-\Pi^{\prime}}.
\end{equation}
The Green's function for the dual bosons can be found from the exact relation
\begin{equation}  \label{ExactW}
\alpha^2 \tilde{X}^ W=W X W+W.
\end{equation}
One can realize that there are two special cases. First, let us take $%
W=V+\Pi^{\prime}-\Lambda$, and use $\alpha=\chi^{-1}$, as in the main body
of the paper. For this choice, using (\ref{W1}, \ref{ExactW}) one can
straightforwardly show that
\begin{equation}  \label{GGprime}
\tilde{X}^{V+\Pi^{\prime}-\Lambda}=\frac{\chi
(V+\Pi^{\prime}-\Lambda)}{\chi^{-1}+\Lambda-V-\Pi^{\prime}}\equiv X - \chi.
\end{equation}
Summing over $k$ and taking Eq.(\ref{DBcondition}) into account we conclude that the
condition (\ref{dual0}) is fulfilled
\begin{equation}
\sum_k \tilde{X}^{V+\Pi^{\prime}-\Lambda}=0.
\end{equation}
It turns out that \textit{the dual propagator vanishes for variables obtained
from the Hubbard-Stratonovich decoupling of the renormalized interaction,
with the hybridization part \textbf{excluded}}, if the condition (\ref{DBcondition})
is fulfilled. For the EDMFT approximation, $\Pi^{\prime}$ is absent and the
condition (\ref{dual0}) holds for the decoupling with $V+\Pi^{\prime}$
used in the present paper.

Now, let us consider the case $W=V+\Pi^{\prime}$. It is suitable to take $%
\alpha=\chi^{-1}+\Lambda$ in this case. We obtain
\begin{equation}  \label{EDMFTGGprime}
\tilde{X}^{V+\Pi^{\prime}}=\frac{(\chi^{-1}+\Lambda)^{-1}(V+\Pi^{%
\prime})}{\chi^{-1}+\Lambda-V-\Pi^{\prime}}\equiv X -
(\chi^{-1}+\Lambda)^{-1}.
\end{equation}
Again taking a summation over $k$ and substituting Eq.(\ref{DBcondition}), we obtain
$\sum_k \tilde{X}^{V+\Pi^{\prime}}=\chi \Lambda \left(\Lambda^{-1}+
\chi\right)^{-1}$. The second line of (\ref{ExactV}) with $%
\alpha=\chi^{-1}+\Lambda$ gives the same answer for $\chi_\phi$, therefore
we obtain
\begin{equation}  \label{ExactWW}
\sum_k \tilde{X}^{V+\Pi^{\prime}}=\chi_\phi.
\end{equation}
\textit{The ``effective-medium'' condition holds
for variables obtained from the Hubbard-Stratonovich decoupling of the
renormalized interaction, with the hybridization part \textbf{included}},
if  condition (\ref{DBcondition}) is fulfilled. For EDMFT, this corresponds to
the decoupling of purely $V \rho^* \rho$ part used in Ref. \cite
{Sun_Kotliar_EDMFT}.

We stress that, strictly speaking, apart from the simple EDMFT case the
conditions (\ref{dual0}), (\ref{dualEM}) are not fulfilled simultaneously with
(\ref{dual0}). This does not cause serious problems for the GW+EDMFT
calculations, because one typically does self-consistent calculations of the
EDMFT problem and then calculates diagram corrections for a pure EDMFT
hybridization. On the other hand, for higher order diagrammatic
approximations the difference can be essential.

Finally, we note that a specific choice of the decoupling procedure is
physically irrelevant. As we have seen, EDMFT can be constructed using $%
W=V-\Lambda$ or $W=V$, and the result for initial ensemble is the same. It
can be checked that any other decoupling will also work; what matters is the
Gaussian approximation for dual variables and the value of the hybridization.
The latter is defined by the condition (\ref{dual0}), that does not include
any dual quantities. We use the theory based on $W=V-\Lambda$ decoupling
because it allows a simple construction of the dual diagrammatic series
without a double counting of the local contributions. This important property
is based on the fact that the EDMFT a dual Green's function has a very simple physical
meaning: according to (\ref{GGprime}), it is just the Green's
function of initial variables with local part subtracted.



\bibliographystyle{elsarticle-num}








{\bf Figures captions:}

Fig.1 Basic building blocks for dual-boson diagram:
fermionic ($\tilde{\mathcal{G}}$) and bosonic ($\tilde{\mathcal{X}}$)
dual propagators as well as $\lambda$ and $\gamma$ vertices.

Fig.2 The bosonic dual self-energy in the ladder
approximation. A triangle represents the $\protect\lambda$ vertex and a square
represents the $\protect\gamma$ vertex.

Fig.3  The bosonic dual self-energy with the renormalized triangle vertex.

Fig.4 The diagrammatic equation for the renormalized triangle vertex.

Fig.5 An example of diagram for the fermionic dual self-energy with the renormalized triangle vertex.

\newpage

\end{document}